\documentclass[final,5p,times,twocolumn]{elsarticle}

\usepackage{amsmath}
\usepackage{graphicx}% Include figure files
\usepackage{xcolor}
 \usepackage{multirow}
 \usepackage{booktabs}
\usepackage{orcidlink}
\usepackage{hyperref}% add hypertext capabilities
\begin{document}

\title{Optical transients from non-explosive double white-dwarf mergers: the case of a central neutron star remnant}

\author[unife,icranet]{M. M. Ridha Fathima \orcidlink{0000-0002-6118-8129} \corref{cor1}}

% \ead{ridhafathima.mohideenmalik@unife.it}

\author[ufes]{Alexandre M.R. Almeida \orcidlink{0000-0003-4227-2250} \corref{cor1}}
% \ead{alexandre.m.almeida@ufes.br}

\author[unife,inafoaab,infnfe]{M. Bulla \orcidlink{0000-0002-8255-5127}}
% \ead{mattia.bulla@unife.it}

\author[ufes]{Jaziel G. Coelho \orcidlink{0000-0001-9386-1042}}
% \ead{jaziel.coelho@ufes.br}

\author[unife,inafoas,infnfe]{C. Guidorzi \orcidlink{0000-0001-6869-0835}} 
% \ead{guidorzi@fe.infn.it}

\author[unife,icranet,icra,icranetferrara,inafaps]{J. A. Rueda \orcidlink{0000-0003-4904-0014}}
% \ead{jorge.rueda@icra.it}

\address[unife]{Dipartimento di Fisica e Scienze della Terra, Universit\`a degli Studi di Ferrara, Via Saragat 1, 44122 Ferrara, Italy}
\address[ufes]{Núcleo de Astrofísica e Cosmologia (Cosmo-Ufes) \& Departamento de Física, Universidade Federal do Espírito Santo, Av. Fernando Ferrari 514,  29075--910, Vitória, ES, Brasil}
\address[icranet]{ICRANet, Piazza della Repubblica 10, 65122 Pescara, Italy}
\address[inafoas]{INAF - Osservatorio di Astrofisica e Scienza dello Spazio di Bologna, Via Piero Gobetti 101, 40129 Bologna, Italy}
\address[infnfe]{INFN - Sezione di Ferrara, Via Saragat 1, 44122 Ferrara, Italy}
\address[inafoaab]{INAF - Osservatorio Astronomico d’Abruzzo, via Mentore Maggini snc, 64100 Teramo, Italy}

\address[icra]{ICRA, Dipartimento di Fisica, Sapienza Università di Roma, Piazzale Aldo Moro 5, I-00185 Roma, Italy}

\address[icranetferrara]{ICRANet-Ferrara, Dipartimento di Fisica e Scienze della Terra, Universit\`a degli Studi di Ferrara, Via Saragat 1, 44122 Ferrara, Italy}

\address[inafaps]{INAF, Istituto di Astrofisica e Planetologia Spaziali, Via Fosso del Cavaliere 100, I-00133 Rome, Italy}

\cortext[cor1]{These authors contributed equally to this work.}

\begin{abstract}

Discoveries of ultra-massive magnetic white dwarfs (WDs) and peculiar pulsars have been proposed to originate in double white dwarf (DWD) mergers. There are three possible post-merger central remnants of non-explosive mergers: 1) a stable sub-Chandrasekhar WD; 2) a rapidly rotating super-Chandrasekhar WD; 3) a neutron star (NS). 
In this work, we explore the thermal transient arising from non-explosive DWD mergers that leave an NS remnant from the prompt collapse of the merged core. The transient is powered by the cooling of the expanding dynamical ejecta, with energy injection from magnetic dipole radiation, which depends on the dipole factor $D = B_d^2/P_0^4$, with $B_d$ and $P_0$ being the surface magnetic field strength and initial rotation period of the newborn NS. We simulate lightcurves in the Legacy Survey of Space and Time (LSST) bands and estimate the horizon and detection rates for these transients across a range of model parameters. We find LSST detection horizons upper limits ranging $30$--$820$ Mpc and corresponding detection rates $10^2$--$10^6$ yr$^{-1}$ for $\log D = 24$--$40$. Accounting for the survey cadence, we find that only configurations with $\log D = 36$--$40$ are detectable within $240$--$760$ Mpc, with detection rates $10^4$--$10^5$ yr$^{-1}$. Combined searches across surveys can compensate for the low cadence and improve the detection rates of fast and less energetic sources. Multi-wavelength campaigns can aid in detecting the spindown radiation at higher energies observable after the optical transient. Observations of these transients will provide direct evidence of the non-explosive DWD mergers, characterise the remnants and progenitor parameters, and the fraction of explosive and non-explosive mergers.
\end{abstract}

% Keywords

\date{\today}

\maketitle

%%%%%%%%%%%%%%%%%%%%%%%%%%%%%%%%%%%%%%%%%%%%%%%%%%%%%%%
%%%%%%%%%%%%%%%%%%%%%%%%%%%%%%%%%%%%%%%%%%%%%%%%%%%%%%%
\section{Introduction}
%%%%%%%%%%%%%%%%%%%%%%%%%%%%%%%%%%%%%%%%%%%%%%%%%%%%%%%
%%%%%%%%%%%%%%%%%%%%%%%%%%%%%%%%%%%%%%%%%%%%%%%%%%%%%%%

Electromagnetic transients, from the radio to the X-rays, originating from non-explosive DWD mergers have been proposed to be detectable with upcoming deep surveys (e.g., \cite{2019JCAP...03..044R, 2023ApJ...958..134S, 2019ApJ...870L..23Y}) and, to be associated with observed sources like the fast-blue optical transients (FBOTs; see, e.g., \cite{2019MNRAS.487.5618L}). 
In this article, we focus on the possible optical emission from the thermal cooling of the dynamical ejecta expelled during the merger process. 
In non-explosive mergers, three post-merger central remnants are possible: a stable sub-Chandrasekhar WD, a rapidly rotating super-Chandrasekhar WD, or a neutron star (NS).
\cite{2023ApJ...958..134S} estimated this emission with the further injection of energy from the fallback accretion onto the central remnant, assuming for the latter a massive, fast-rotating, stable sub-Chandrasekhar WD. 
The predicted transient has optical peak luminosities of the order of $10^{41}$ erg s$^{-1}$ at $\sim 1$--$1.5$ days post-merger. Such a transient lies in the gap between supernovae and novae (See Figure 3 in \cite{2023ApJ...958..134S}).
Detection of this class of transients requires deep, high-cadence optical surveys, achievable with deep and large-scale surveys like the Zwicky Transient Facility (ZTF) at Palomar Observatory \citep{2019PASP..131a8002B} and the Vera C. Rubin Observatory’s Legacy Survey of Space and Time (LSST; \cite{2019ApJ...873..111I}). With the estimated rate of DWD mergers within Hubble time, \cite{2023ApJ...958..134S} found the optical transient properties to be consistent with their lack of detection by the ZTF to date, and set an upper limit of $10^3$ detections per year for the LSST.  Our aim here is to improve both the modelling and the estimation of the detection rates of this class of optical transients. We broaden the scope of this series of works to include an additional scenario: a central neutron star (NS), whose magnetic braking provides energy injection. We consider the effects of thermalisation and energy leakage. Finally, we estimate detection rates for the LSST in each photometric band, accounting for the survey cadence. We find that, although potentially less abundant, a central magnetised NS will produce powerful transients with higher detection rates than those from fallback onto a central WD.

The paper is structured as follows: Section \ref{sec:heating} models the energy injection into the ejecta by the remnant; Section \ref{sec:transient} presents the results of the simulations of a transient powered by fast spinning, highly magnetised NS; Section \ref{sec:detection} provides the detection rate of such events. Finally, Section \ref{sec:discussion} summarises and discusses the implications of the work.

%%%%%%%%%%%%%%%%%%%%%%%%%%%%%%%%%%%%%%%%%%%%%%%%%%%
%%%%%%%%%%%%%%%%%%%%%%%%%%%%%%%%%%%%%%%%%%%%%%%%%%%
\section{Heating and Cooling of the Dynamical Ejecta}
\label{sec:heating}
% Section goal: Describing the model
%%%%%%%%%%%%%%%%%%%%%%%%%%%%%%%%%%%%%%%%%%%%%%%%%%%
%%%%%%%%%%%%%%%%%%%%%%%%%%%%%%%%%%%%%%%%%%%%%%%%%%%

% Para goal: Aim of the section and ejecta profile
Here, we describe the semi-analytical model of the emission from the expansion and cooling of the dynamical ejecta, heated by the central merger remnant. 
The ejecta is modelled as spherically symmetric, non-interacting shells surrounding the remnant, following a power-law density profile, and expanding self-similarly.
We refer to \ref{app:ejecta} for further details of the ejecta properties and modelling.

%%%%%%%%%%%%%%%%%%%%%%%%%%%%%%%%%%%%%%%%%%%%%%%%%%%
\subsection{The post-merger central remnant}
%%%%%%%%%%%%%%%%%%%%%%%%%%%%%%%%%%%%%%%%%%%%%%%%%%%

The diversity of these outcomes depends upon the masses and compositions of the WD components. 
The DWD merger may trigger a thermonuclear explosion, such as a Type Ia supernova (SN Ia \citep{1984ApJS...54..335I, 1985ASSL..113....1P}) or a Calcium-strong transient (CaST \citep{2010Natur.465..322P}). 
However, if the merger does not achieve the central density and temperature to ignite the unstable carbon fusion, the merger will leave a stable or metastable post-merger central remnant \citep{1984ApJ...277..355W, 2012ApJ...748...35S, 2014MNRAS.438...14D, 2018ApJ...857..134B}. 
From these non-explosive scenarios \citep{1985ASSL..113....1P}, there are three final fates for the central remnant:
\begin{enumerate}
    \item[1)] A massive, fast-rotating, and highly magnetized stable sub-Chandrasekhar WD (see, e.g., \cite{2012PASJ...64...56M, 2013ApJ...772L..24R, 2017MNRAS.465.4434C, 2020ApJ...895...26B, 2020NatAs...4..663H, 2025arXiv250713850C} for their possible high-energy emission, and \cite{2018MNRAS.479L.113K, 2021Natur.595...39C, 2021ApJ...923L...6K, 2022ApJ...941...28S, 2025ApJ...994...12W}, for their recent detection and discussion on their likely DWD merger origin);
    \item[2)] A metastable super-Chandrasekhar WD that avoids prompt gravitational collapse due to its rapid rotation (see, e.g., \cite{2018ApJ...857..134B, 2019MNRAS.487..812B});
    \item[3)]  An NS from prompt collapse shortly after the merger (see, e.g., \cite{1985A&A...150L..21S, 1985ApJ...297..531N})\footnote{Delayed collapse into an NS of massive, highly magnetized WDs occurs on a much longer timescale than the time of occurrence of the post-merger transient analysed here (see, e.g., \cite{2019MNRAS.487..812B}).}.
\end{enumerate}

Since NSs are characterised by higher magnetic fields and are faster rotators than WDs, such a central remnant should produce a much more powerful energy injection into the expanding ejecta than in the case of a post-merger WD, hence leading to a more luminous and easier-to-detect transient. We will return to this point in Section \ref{sec:transient}.

%%%%%%%%%%%%%%%%%%%%%%%%%%%%%%%%%%%%%%%%%%%%%%%%%%%
\subsection{Energy injected into the ejecta}
%%%%%%%%%%%%%%%%%%%%%%%%%%%%%%%%%%%%%%%%%%%%%%%%%%%
% Para goal: Energy sources

The initial energy from the merger thermalises the ejecta, and the presence of a remnant further injects energy into it. One possible mechanism by which the presence of remnant injects additional energy is the fallback accretion. This is the luminosity emitted by the material initially released during the merger, which remains gravitationally bound to the remnant and falls back, releasing energy \citep{2007MNRAS.376L..48R}. The fallback luminosity can be modelled as
\begin{equation}\label{eq:Hfb}
    H_{\rm fb}(t) = \frac{H_{\rm fb,0}}{(1 + t/t_{\rm fb,c})^{\lambda}}\,.
\end{equation}
Based on numerical simulations of DWD mergers (see, e.g., \cite{2019JCAP...03..044R}, and references therein),  \cite{2023ApJ...958..134S} used Eq. (\ref{eq:Hfb}) with $H_{\rm fb,0}=10^{46}$ erg s$^{-1}$, $\lambda=1.3$ and $t_{\rm fb,c}=100$ s as model parameters for a WD remnant (see \ref{app:comparison}). In the case of an NS remnant, the magnetic pressure from the highly magnetic field can slow the process, reducing the fallback. Under those conditions, the braking radiation becomes the main energy source for the ejecta. The spindown luminosity by magnetic dipole braking is given by
\begin{equation}\label{eq:Hsd}
    H_{\rm sd}(t) = \frac{H_{\rm sd,0}}{(1 + t/t_{\rm sd})^2}, \quad H_{\rm sd,0}=\frac{32\pi^4 B_d^2 R^6}{3 c^3P_0^4}\,,
\end{equation}
where $H_{\rm sd,0}$ is the initial spindown power of a dipole with magnetic strength $B_d$ at the pole, an initial rotation period $P_0$, and radius $R$. The quantity $t_{\rm sd}$ is the characteristic timescale for spindown

\begin{equation}\label{eq:tsd}
    t_{\rm sd}=\frac{3Ic^3P_0^2}{16\pi^2B_d^2R^6}\,,
\end{equation}
where $I$ is the stellar moment of inertia.

% Para goal: Efficiency of injection
When the compact remnant is stable, the spindown timescale is significantly longer than the timescale of the observed transient.  For typical NS values, such as $B_d\sim 10^{12}$ G, $R\sim10^6$ cm, $P_0\sim 1$ ms, and $I\sim10^{45}$ g cm$^2$, we have $t_{\rm sd} \approx 16$ years. %dipole factor = 36
Thus, rather than the spindown timescale, the absorption and thermalisation of the ejecta determine the transient timescale.

% Para goal: Define absorption efficiency
Initially, when the ejecta is optically thick, the optical depth is $\tau_0 \gg 1$, and the injected power from the central remnant is completely trapped. As the ejecta expands (see Eq. \ref{eq:ri} in \ref{app:ejecta}) and becomes optically thin, i.e., $\tau_0 \sim 1$, the injected luminosity leaks out of the ejecta without interaction. To account for this trapping and leakage, we define the absorption parameter
\begin{equation}\label{eq:eta}
    \eta = 1-e^{-\tau_0}\,.
\end{equation}
From the definition of $\eta$ given by Eq. (\ref{eq:eta}), the effective heating rate of the ejecta by the injected power $H_{\rm inj}$ can be written as
\begin{equation}\label{eq:Heff}
    H_{\rm eff} = \eta \cdot H_{\rm inj}\,,
\end{equation}
and the energy that escapes per unit time is
\begin{equation}\label{eq:Lleak}
    L_{\rm leak} = H_{\rm inj}- H_{\rm eff} = e^{-\tau_0} H_{\rm inj}\,,
\end{equation}
where $H_{\rm inj} = H_{\rm sd}+H_{\rm fb}$ if both fallback and magnetic braking, given by Eqs. (\ref{eq:Hfb}) and (\ref{eq:Hsd}), are considered, and we have used Eqs. (\ref{eq:eta}) and (\ref{eq:Heff}).

% Para goal: Introduce heating efficiency
We divide the ejecta into $N$ layers (see \ref{app:ejecta}). The mass of the $i$-th layer is determined by 
\begin{equation}\label{eq:mi}
    m_i = \int_{r_{i}}^{r_{i+1}} \rho(r)\, 4\pi r^2 dr\,,
\end{equation}
where $\rho(r)$ is the ejecta density profile (see Eqs. \ref{eq:rho} and \ref{eq:rho0} in \ref{app:ejecta}). As the ejecta density falls with distance following a power-law, the inner layers are denser than the outer layers and are therefore expected to experience higher effective heating. In the absence of a full radiative transport calculation, we model this effect by assuming a constant effective heating per unit mass throughout the ejecta. Under this assumption, the heating of the $i$-th layer is given by
\begin{equation}
    H_{{\rm eff},i} = \frac{m_i}{m_{\rm ej}} H_{\rm eff}\,.
\end{equation}
where $m_i$ is given by Eq. (\ref{eq:mi}), and $m_{\rm ej}$ is the total ejecta mass.
 
%%%%%%%%%%%%%%%%%%%%%%%%%%%%%%%%%%%%%%%%%%%%%%%%%%%%%%%%%
\subsection{Energy radiated by the ejecta}

% Para goal: Energy equation
We now turn to describe the evolution of the ejecta and its radiation. For this task, we solve the energy conservation equation of each layer of the ejecta
\begin{equation}\label{eq:energyconservation}
    \dot{E}_i = -P_i \dot{V}_i - L_{{\rm cool},i} + H_{{\rm eff},i} \,,
\end{equation}
where $E_i$ is the internal energy, $P_i$ is the pressure, and $V_i$ is the volume of each shell. The radiative cooling term, $L_{{\rm cool},i}$, represents the radiated emission from each shell of ejecta due to photon diffusion. It is derived under the assumption that photons travel through the expanding medium at a rate determined by the local optical depth $\tau_i$, i.e.,
\begin{equation*}
    L_{{\rm cool},i} = \frac{cE_i e^{-\tau_i}}{r_i}\,,
\end{equation*}
where $\tau_i$ is given in Eq. (\ref{eqn:tau_i}) in \ref{app:ejecta}.

To close the system of equations, a radiation-dominated equation of state is implemented: $ E_i=3P_iV_i$. These equations are integrated forward in time for each shell. The total bolometric luminosity, from the system of ejecta and remnant after a DWD merger, is the sum of the cooling luminosity of the ejecta and the leaked luminosity from the remnant:
\begin{equation}\label{eq:Lbol}
       L_{\rm bol}(t) = L_{\rm rad}(t)+L_{\rm leak}(t)\,,
\end{equation}
where $L_{\rm rad} = \sum_{i} L_{{\rm cool},i}(t)$ is the total ejecta thermal radiation.

%Para goal: spectral evolution
The expanding ejecta is a blackbody emitter of effective temperature $T_{\rm eff}(t)$, and photospheric radius $R_{\rm ph}(t)$ giving the radial depth from which photons can escape freely, which is given by the condition $\tau_0(R_{\rm ph}, t)\approx 1$ (see Eq. \ref{eqn:photo} in \ref{app:ejecta}). After the ejecta becomes transparent, the photosphere is set to evolve as the innermost radius of the ejecta. We obtain the evolution of the effective blackbody temperature of the ejecta, $T_{\rm eff}(t)$, from the Stefan-Boltzmann law
\begin{equation}\label{eq:Lbb}
    L_{\rm rad}(t) = 4\pi R_{\rm ph}^2(t) \sigma T_{\rm eff}^4(t),
\end{equation}
where $\sigma$ is the Stefan-Boltzmann constant. The luminosity emitted in a given photometric band is obtained by integrating the Planck function $B_\nu(T_{\rm eff})$ over the corresponding frequency range ($\nu_1$--$\nu_2$), weighted by the filter transmission function of the detector in that band, $S_{\nu,\rm band}$, i.e.,
\begin{equation}\label{eq:Lband}
    L_{\rm band}(t) = \int_{\nu_1}^{\nu_2} 4\pi R_{\rm ph}^2(t) B_\nu(T_{\rm eff}) \cdot S_{\nu,\text{band}} \, d\nu \,.
\end{equation}
This time-dependent luminosity in the observed band can then be used to compute the absolute magnitude.

%%%%%%%%%%%%%%%%%%%%%%%%%%%%%%%%%%%
\section{The post-merger transient}
\label{sec:transient}
% Section goal: Simulation results
%%%%%%%%%%%%%%%%%%%%%%%%%%%%%%%%%%%

% Para goal: Describe the bolometric
\begin{figure}[ht]
    \centering
    \includegraphics[width=\linewidth,clip]{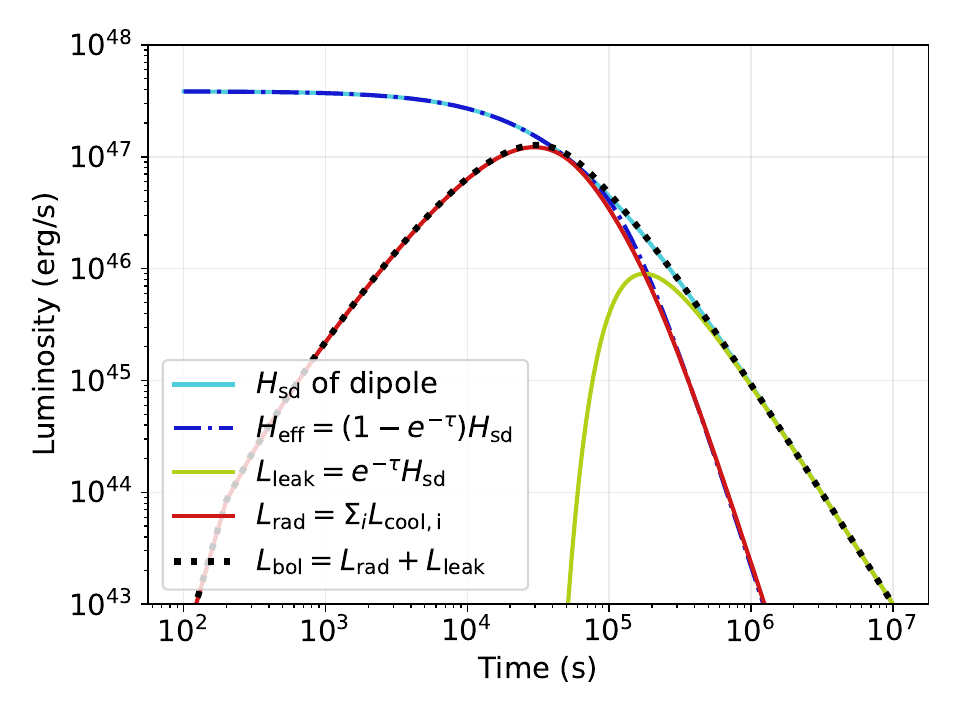}
    \caption{Time evolution of heating by the source ($H_{\rm inj} = H_{\rm sd}$, light-blue solid line), the effective injection into the ejecta ($H_{\rm eff}$, dark-blue dot-dashed line), and the thermal radiation emitted ($L_{\rm rad}$, red solid line). The bolometric luminosity ($L_{\rm bol}$) is the sum of the radiated luminosity from the cooling of the ejecta ($L_{\rm rad}$, black dotted line) and the leaked luminosity ($L_{\rm leak}$, green solid line). The source considered here is a highly magnetic ($B_d = 10^{14}$ G) and rapidly rotating ($P_0 = 1$ ms) NS.}
    \label{fig:components}
\end{figure}

We now analyse the properties of the transient arising from a DWD merger that leads to a central NS remnant. We recall that we assume the ejecta is irradiated only by the NS dipole braking power, neglecting fallback, which is expected to be reduced by the magnetic pressure provided by the high magnetic fields considered here. Indeed, the Alfv\'en radius $r_A = [B_d R^6/(\dot M \sqrt{2 G M})]^{2/7}$ for accretion rates lower than the Eddington limit is well above the NS surface for all magnetic fields of interest. Figure \ref{fig:components} shows the bolometric lightcurve of a transient powered by a highly magnetic ($B_d = 10^{14}$ G) and rapidly rotating ($P_0 = 1$ ms) NS. These NS parameters are chosen to illustrate the behaviour of the various components of the bolometric luminosity for a strong energy injection. However, they may not represent the most plausible outcome of a DWD merger. The knowledge of the latter requires prior knowledge of the distribution of central remnant parameters inferred from the DWD population, and the detection rate of the transients analysed here could be a powerful tool for their characterisation.

Examining the behaviour of the components of the bolometric ($L_{\rm rad}$ and $L_{\rm leak}$), the timescale and peak are mainly determined by the balance between the energy injection from the remnant and the time-dependent trapping efficiency $\eta(t)$. For instance, the luminosity that leaks peaks when the ejecta becomes fully transparent, in this example, at $t \approx 1.5 \times 10^5$\,s.

% Para goal: Describe the cooling for various cases

\begin{figure}[ht]
    \centering
    \includegraphics[width=\linewidth,clip]{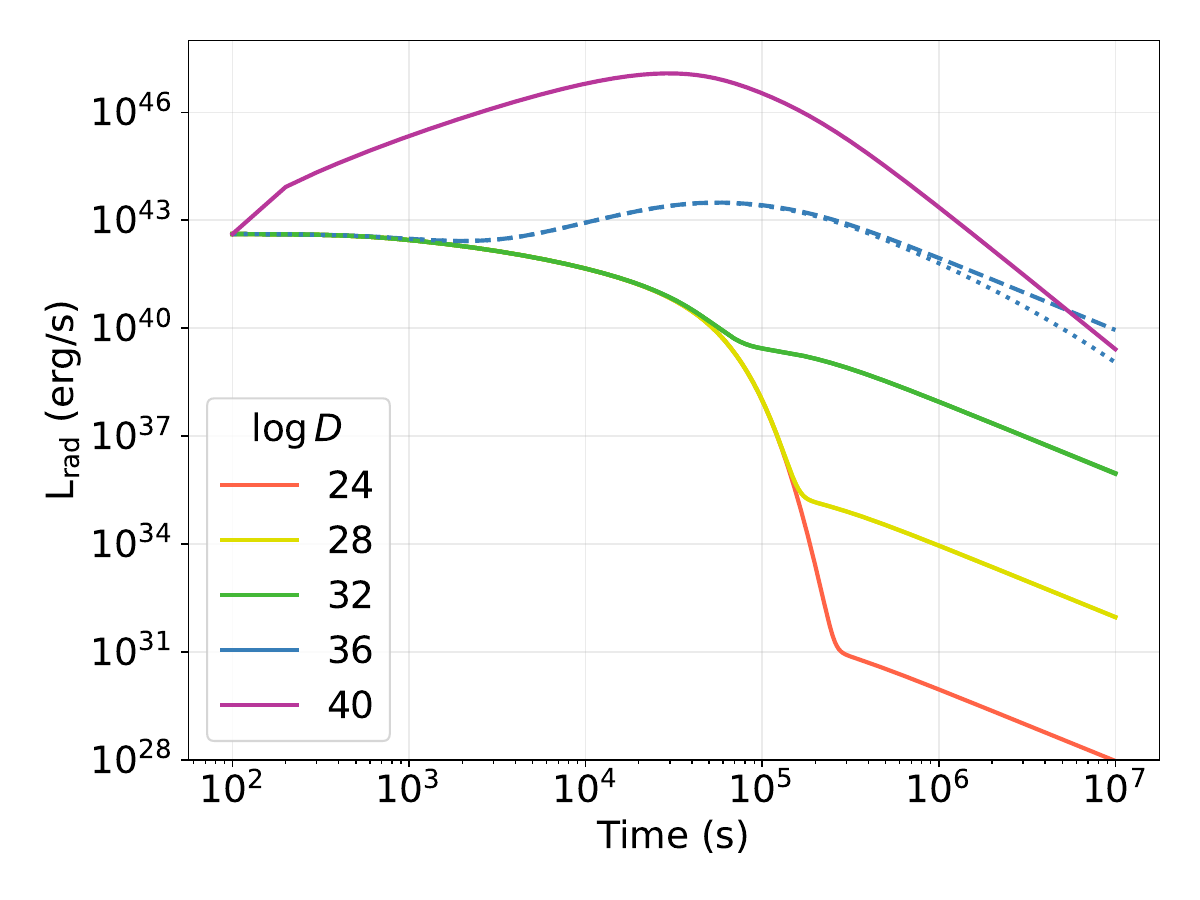}
    \caption{Isotropic luminosity radiated by the ejecta powered by NS remnant with dipole factors \(\log D=24\)--40, corresponding to configurations with magnetic fields \(B_d=10^{10}\)--\(10^{14}\) G and initial rotation periods \(P_0 = 1\)--\(100\) ms. The change in the slope of the curves at late times indicates the transition to the adiabatic cooling regime. The dotted and dashed blue curves are configurations with the same dipole factor, $\log D = 36$, but with different magnetic field and rotation period, \{$B_d = 10^{14}$ G, $P_0 = 10$ ms\} and \{$B_d = 10^{12}$ G, $P_0 = 1 $ ms\}.}
    \label{fig:dipole-factors}
\end{figure}

Since the ejecta is powered by spindown, the radiated luminosity varies with the magnetic field and period. To compare the diversity of the transients for different values of the magnetic field and period, we introduce the dipole factor, defined as $D \equiv B_d^2/P_0^4$. Figure \ref{fig:dipole-factors} shows the results for configurations with $P_0 = 1$--$100$ ms and $B_d = 10^{10}$--$10^{14}$ G, which translates into a dipole factor in the range $\log D = 24$--$40$ (see Figure \ref{fig:components} for the case of $\log D = 40$). Generally, the higher the factor, the brighter the transient.

\begin{figure}[ht]
    \centering
    \includegraphics[width=1.\linewidth,clip]{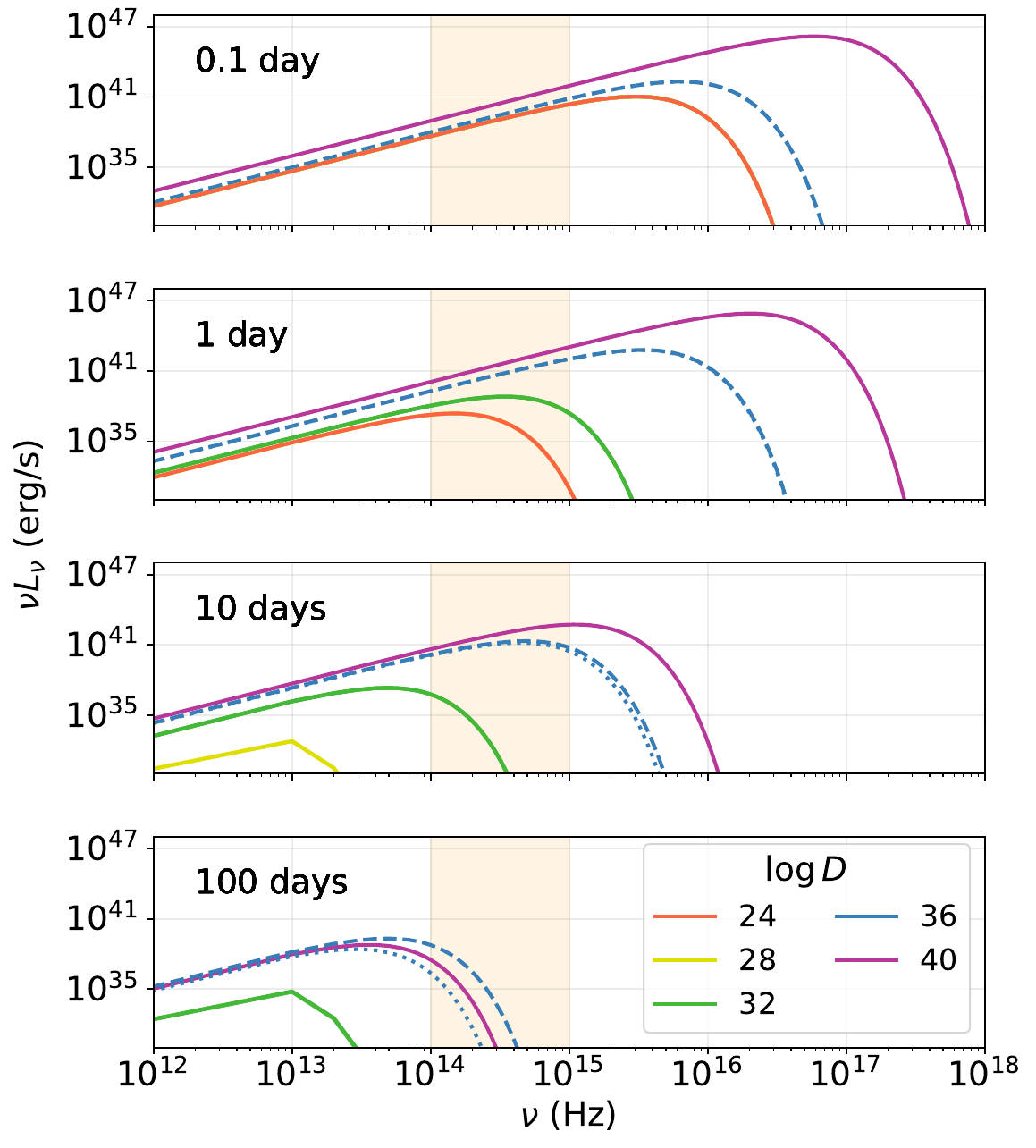}
    \caption{Temporal evolution of the thermal spectra of the ejecta heated by the NS remnant of a DWD merger. The optical range is marked by the shaded region between $10^{14}$--$10^{15}$ Hz. The dipole factor characterises the strength of the injected dipole radiation. In the upper plot, lightcurves corresponding to $\log D = 24, 28$ and $32$ (red, yellow and green, respectively) overlap. Even at $t \sim 1$ day, $\log D = 24$ and $28$ are overlapping, indicating that low dipole factors lead to similar early-time behaviour. The dotted and dashed blue lightcurves correspond to configurations with \{$B_d = 10^{14}$ G, $P_0 = 10$ ms\} and \{$B_d = 10^{12}$ G, $P_0 = 1 $ ms\}, which show how the behaviour of the transient does not depend on the dipole factor alone (see text for details).}
    \label{fig:spectra}
\end{figure}

Figure \ref{fig:spectra} shows the spectral evolution on selected post-merger times in the range $0.1$--$100$ days. It is noted that for lower dipole factors, the spectral peaks in the optical wavelengths ($300$--$700$ nm or $10^{14}$--$10^{15}$ Hz) occur within a day of the trigger and then decay rapidly. The configurations with high dipole factors peak brighter in the optical at later times and remain peaked 100 days post-merger.

A given dipole factor can represent different configurations with different magnetic fields and rotation periods, and these configurations can have different behaviour. This is the case shown in Figure \ref{fig:spectra} of the dipole factor $\log D = 36$, which shows a late-time deviation for two configurations: \{$B_d = 10^{14}$ G, $P_0 = 10$ ms\} (dotted blue) fades faster than when \{$B_d = 10^{12}$ G, $P_0 = 1 $ ms\} (dashed blue). The reason is that, as shown by Eq. (\ref{eq:tsd}), the spindown timescale scales with $1/(D P_0^2)$, so it is shorter for the former than for the latter.

In \ref{app:comparison}, we have also applied the present improved treatment of the ejecta and its transparency to the case of a post-merger, massive, fast rotating, highly magnetized WD, which is helpful to compare with previous results \citep{2023ApJ...958..134S}.

%%%%%%%%%%%%%%%%%%%%%%%%%%%%%%%%%%%%%%%%%%%%%%%%%%%%%%%%%
\section{Detection capabilities with large-scale surveys}
\label{sec:detection}
% Section goal: lightcurves, detection horizon and rate
%%%%%%%%%%%%%%%%%%%%%%%%%%%%%%%%%%%%%%%%%%%%%%%%%%%%%%%%%

\begin{figure*}[ht]
    \centering
    \includegraphics[scale=0.38]{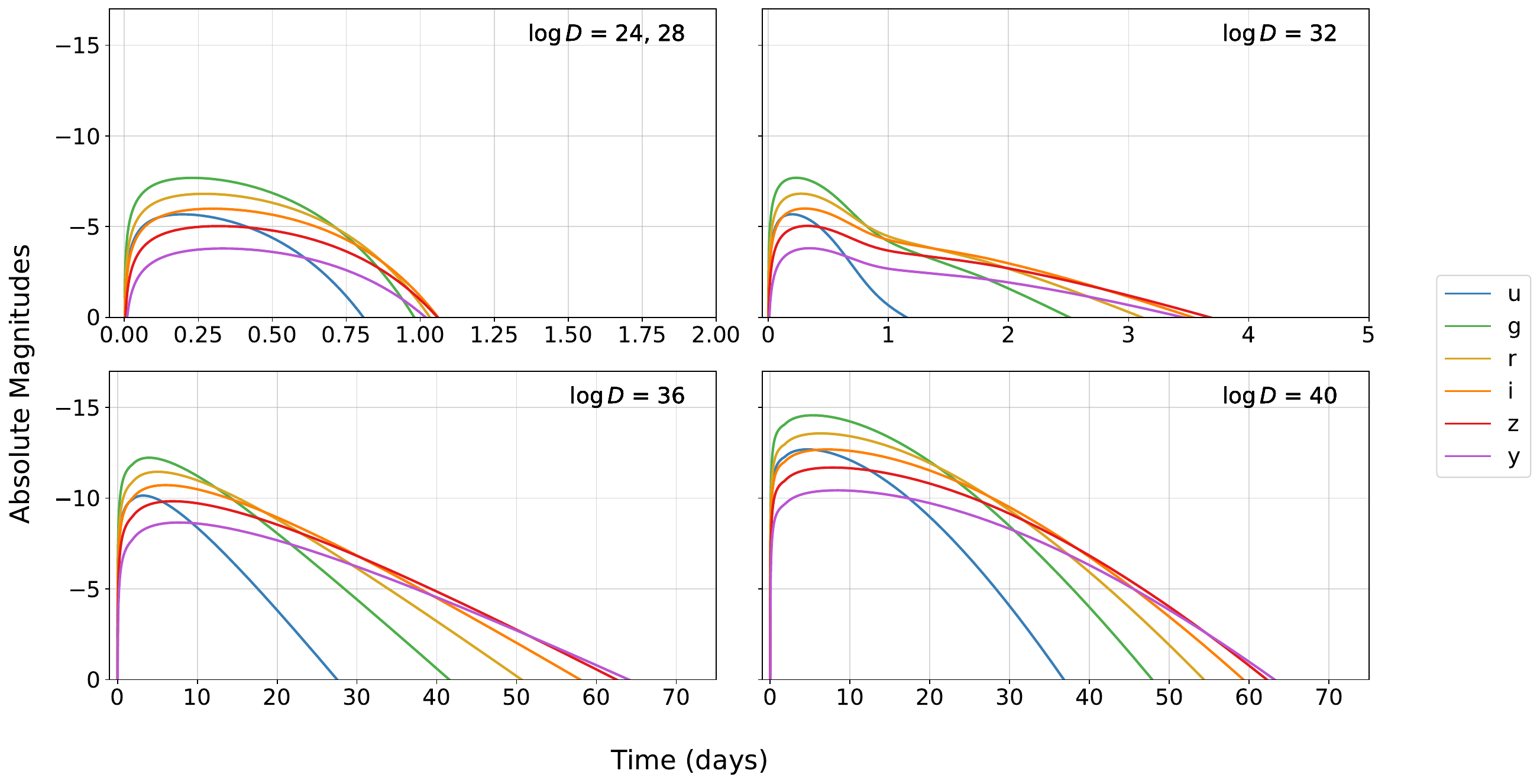}
    \caption{A grid of DWD merger transients powered by NS  remnant with varying magnetic fields and rotations. The lightcurves, in absolute magnitudes, are simulated in the LSST \textit{ugrizy} bands. The configurations with $\log D = 24$ and $28$ produce similar lightcurves and are therefore presented together. The configurations with $\log D = 36$ show a slight variation in the decay time. Here, we show the fastest case.} 
    \label{fig:lightcurve}
\end{figure*}

% Para goal: Describe the conversion
We convert the luminosities to absolute magnitudes as usual,
\begin{equation}\label{eqn:conversion}
    M_{\rm band}(t,z) = -2.5 \log \left( \frac{L_{\rm band}(t)}{4\pi d_L^2 \Delta\nu} \right) - 48.6\,,
\end{equation}
where the luminosity distance is $d_L(z) = 10$ pc, and the effective bandwidth is $\Delta\nu = \int_{\nu_1}^{\nu_2} S_\nu d\nu$ is the effective bandwidth. Figure \ref{fig:lightcurve} shows the evolution of the transient in the LSST $ugrizy$ bands, for the dipole factors of Figure \ref{fig:dipole-factors}. The duration of the transient varies widely. The brightest lasts for months, whereas the faintest decays rapidly within a day of peak. The transients are blue at early times, but their colour evolves (reddening) during the decay.

% Para goal: Describe the horizon
Usually, the detection horizon of an astronomical source for a given telescope is obtained considering the peak magnitude of the source, $M_{\rm peak}$, as
\begin{equation}\label{eq:dmax}
    \log d_{\rm max} {\rm [pc]} = 0.2 (m_{\rm lim} - M_{\rm peak} + 5).
\end{equation}
$m_{\rm lim}$  is the limiting magnitude for the particular photometric band of a survey. We use the $5\sigma$ limits from \cite{2022ApJS..258....1B} for the LSST and \cite{2019PASP..131a8002B} for the ZTF.

However, to characterise a transient and associate it with a progenitor, more than one observation is necessary. By considering the brightness after the peak at a given survey cadence $t_{\rm cad}$, $M_{\rm cad} \equiv M(t_{\rm cad})$, we obtain more realistic detection horizons and rates. In our estimates, we assume $t_{\rm cad} = 2$ and $3$ days from the starting time of observation ($t_{i, \rm obs}$) for the LSST and the ZTF. Thus, we define the horizon distance up to which the source can be detected with such a cadence time, $d_{\rm cad}$, as
\begin{equation}\label{eq:dcad}
    \log d_{\rm cad} {\rm [pc]} = 0.2 (m_{\rm lim} - M_{\rm cad} + 5).
\end{equation}

% Para goal: Describe the detection rat
For a transient rate $\mathcal{R}_{*}$ (number of events per unit time per unit volume), the detection rate of an event by a telescope is
\begin{equation}\label{eq:Ndet}
    \mathcal{N} (d < d_\text{max/cad}) = \mathcal{R_{*}} \times \frac{4\pi}{3}d_\text{max/cad}^3 \times f_\text{sky}\,,
\end{equation}
where $f_\text{sky}$ is the sky footprint of the survey, given by the ratio of the area covered by the survey during one cadence to the full sky area.
The ZTF survey covers $15,000 \deg^2$ of the sky in 3 days giving $f_\text{sky} = 0.35$ \citep{2020ApJ...904...35P}. 
The predicted sky coverage of the LSST using the Wide, Fast, Deep survey strategy for transients is $18,000 \deg^2$ in 3 days, producing a sky footprint of $f_\text{sky} = 0.44$ \citep{2023PASP..135j5002H}. 
In Eqs. (\ref{eq:dmax}) and (\ref{eq:dcad}), we adopt the peak time of the lightcurve as the first observation, i.e., $t_{i, \rm obs} = t_{\rm peak}$, which gives the maximum detection rate (see \ref{app:impact}). Other factors that affect the detection rate beyond observation time include Galactic extinction, environment (offset from the Galactic plane), and requirements for different observation strategies and search algorithms, which are not considered in this work.

Since the occurrence fractions of the various outcomes of a DWD merger are not yet known, a back-of-the-envelope calculation is performed as follows. 
As discussed in Section \ref{sec:heating}, DWD mergers can be explosive or non-explosive. Thus, $\mathcal{R}_\text{DWD} \equiv \mathcal{R}_\text{rem} + \mathcal{R}_\text{SNIa, DD}$, where $\mathcal{R}_\text{rem} = k_\text{rem} \mathcal{R}_\text{DWD}$ is the rate of non-explosive mergers, i.e., leaving a central remnant, with $0 < k_\text{rem} <1$, and $\mathcal{R}_\text{SNIa, DD}$ is the rate of SNIa originating from DWD mergers (i.e., double-degenerate channel). 
Taking into consideration that a fraction of SNIa can originate from single degenerate channel, we write $\mathcal{R}_\text{SNIa, DD} = l_\text{DD} \mathcal{R}_\text{SNIa}$, where $l_\text{DD}\leq 1$. Thus, the fraction of SN Ia produced by the single-degenerate channel is $1-l_\text{DD}$. We can now write the DWD merger rate as
\begin{equation}\label{eq:krem}
    \mathcal{R}_\text{DWD} = k_\text{rem} \mathcal{R}_\text{DWD} + l_\text{DD} \mathcal{R}_\text{SNIa}\,.
\end{equation}
The total DWD merger rate within a Hubble time has been estimated to be $\mathcal{R}_{\rm DWD}\sim(5\text{--}7)\times 10^5 \,\mathrm{Gpc}^{-3}\mathrm{yr}^{-1}$  \citep{2024RNAAS...8..323M,2018MNRAS.476.2584M}. Regarding the SN Ia rate, we use in our estimates the locally observed value, $\mathcal{R}_\text{SNIa}\sim(2.1$--$2.6)\times 10^4$ Gpc$^{-3}$ yr$^{-1}$ \citep{2020ApJ...904...35P}. 
We assume local cosmic rates for the DWD merger and SNIa rates, as the maximum horizon we obtain of these events (see below) is about $1$ Gpc (see Table \ref{tab:rates_NS_D_g_LSST_ZTF}), which does not exceed $z \sim 0.2$.

If $k_{\rm rem} = k_{\rm NS} + k_{\rm WD}$, with $k_{\rm NS}$ and $k_{\rm WD}$ being the fractions of non-explosive mergers that produce central NS and WD remnant, Eq. (\ref{eq:krem}) can be rewritten as
\begin{equation}
    k_\text{NS} = 1 - k_\text{WD} - l_\text{DD} \frac{\mathcal{R}_\text{SNIa}}{\mathcal{R}_\text{DWD}}\,.
\end{equation}
Thus, at fixed $k_{\rm WD}$, a lower limit to $k_{\rm NS}$ is obtained for $l_\text{DD} = 1$, i.e., assuming all SN Ia originate from DWD mergers. With this assumption, $k_{\rm NS} + k_{\rm WD}$ is in the range $\sim 0.95$--$0.97$, for the above estimates of $\mathcal{R}_\text{SNIa}$ and $\mathcal{R}_\text{DWD}$. These values can be adjusted for various SNIa and DWD merger models. 
Therefore, we write the rate of transients associated with a central NS (WD) remnant as
\begin{equation}\label{eq:Rwdns}
    \mathcal{R}_{*} = k_\text{NS/WD} \cdot \mathcal{R}_\text{DWD}\,.
\end{equation}
Below, to provide an upper bound on the detection rate of the optical transient of non-explosive mergers leaving an NS, we take $k_{\rm WD} = 0$ and the central values of the rates, i.e., $k_{\rm NS} = 0.96$. 
% Below, to provide an upper bound on the detection rate of the optical transient of non-explosive mergers leaving a NS, we will use the maximum value of $k_{\rm NS}$ using $k_{\rm WD} = 0$, i.e., $k_{\rm NS} = 0.97$. 
% another assumption: all NS formation produces the transient

Table \ref{tab:rates_NS_14_-3_LSST} presents the peak magnitude and time, as well as the detection horizons from Eqs. (\ref{eq:dmax}) and (\ref{eq:dcad}), and the respective upper limits of the detection rate for an NS-powered transient in each of the LSST bands, which are obtained from Eqs. (\ref{eq:Ndet}) and (\ref{eq:Rwdns}). The $g$ and $r$ bands present the highest detection rates. Table \ref{tab:rates_NS_D_g_LSST_ZTF} shows how the detection rate in the $g$ band increases as the dipole factor increases.

The detection rates for a transient powered by fallback accretion onto a WD remnant (see Table \ref{tab:rates_WD_LSST_ZTF_resub} in \ref{app:comparison}) are significantly lower than the case of an NS remnant. For instance, for the LSST \(g\) band, the detection rate accounting for 3-days detector cadence is \(\sim 30\) yr\(^{-1}\), whereas for an NS  with \(\log D = 40\) (see Table \ref{tab:rates_NS_14_-3_LSST}), it reaches \(\sim 1.0 \times 10^6\) yr\(^{-1}\). This contrast indicates that, although less frequent, mergers that yield a highly magnetized, fast-rotating NS remnant are far more promising candidates for detection by wide-field surveys such as LSST. The \ref{app:comparison} shows our results for the g, r, and i bands of LSST and ZTF in the case of a WD remnant powered by fallback accretion, which can be readily compared with previous results by \cite{2023ApJ...958..134S}. The detector cadence is the most critical factor in reducing the detection rate. 

%%%%%%%%%%%%%%%%%%%%%%%%%%%%%%%
\section{Summary \& Discussion}
\label{sec:discussion}
%%%%%%%%%%%%%%%%%%%%%%%%%%%%%%%
% Section goal: Summary of the work. Future work

% Para-goal: Our work and assumptions and near-future work - I. Modelling
We have considered the case of DWD mergers forming an NS remnant. We have simulated optical transients from the expanding dynamical ejecta, powered by the spindown luminosity of the NS remnant. The heating by the remnant thermalises the ejecta, and the blackbody cooling gives the observed optical radiation. We have introduced a leakage in the energy injection onto the ejecta. Similar to spindown, fallback accretion onto the remnant can power the transient, as considered for the case of a WD remnant by \cite{2023ApJ...958..134S} (see \ref{app:comparison}). The amount of fallback must be estimated, accounting for the magnetic pressure from the strong field of a central NS remnant.

% Para-goal: Our work and assumptions and near-future work - I. Detection rates
We use the dipole factor ($D$) to compare transient behaviour. The higher the dipole factor, the brighter and longer the transient. We find that $g$ and $r$ are the optimal photometric bands for detection, and estimate an upper limit on the LSST detection rate in the range $10^2$--$10^6$ yr$^{-1}$, depending on the energetics of spindown injection by the NS. This limit accounts for the transient decay over the survey cadence. The detection rate increases with the dipole factor. For configurations with low values, e.g., $\log D = 28$--$32$, transient detection requires high cadence and deep surveys. The cadence of surveys like the ZTF and the LSST is $2$--$3$ days, which is insufficient for detecting and characterising transients lasting less than a day, e.g., $\log D < 32$ (see Figure \ref{fig:lightcurve}). Searches for and identification of these transients could be enhanced by dedicated detection pipelines and by observing additional predicted characteristic features. For instance, we are implementing a radiative transfer model to characterise the transient spectral signature (Das et al., in prep.). 

It is also worth noting that at early times, the thermal transient peaks in higher-energy bands, e.g., the UV band, relevant for future missions dedicated to the UV transient sky, such as ULTRASAT \citep{2024ApJ...964...74S}, QUVIK \citep{2024SSRv..220...11W} and UVEX \citep{2021arXiv211115608K}, which are expected to fly in 2027, 2028, and 2030, respectively.

% Para goal: If we detect a transient, what can we tell about the progenitor
Observations of these optical-UV transients will provide direct evidence of DWD mergers. They can be used to characterise the remnants and the progenitor parameters, as well as to constrain the fraction of explosive and non-explosive mergers. Since the luminosity and evolution of the associated lightcurve are shaped by the properties of the central remnant and the ejecta, we can estimate the dipole factor ($D$) and mass of ejecta ($m_{\rm ej}$). Modelling the spectrum of the leakage radiation can aid multi-wavelength associations at late times following the optical transient. For instance, observations at higher energies, e.g., in X-rays, could pinpoint the luminosity that leaks, which constrains the spindown timescale ($t_{\rm sd}$). Information on $D$ and $t_{\rm sd}$ can help us infer the critical conditions for collapse using conservation of magnetic flux and period. Since the amount of ejecta depends on the mass ratio of the DWDs and the total mass \citep{2014MNRAS.438...14D}, constraining $m_{\rm ej}$ will help us constrain the DWD progenitor population.

% Para-goal: The future
Furthermore, gravitational waves from DWD binaries are expected to be detected by space-based facilities such as the Laser Interferometer Space Antenna - \textit{LISA} \citep{2017arXiv170200786A}, \textit{TianQin} \citep{2016CQGra..33c5010L}, and \textit{Taiji} \citep{2017NSRev...4..685H}, among which electromagnetic bright binaries peak at 1 kpc \citep{2025A&A...701L...6P}. By combining electromagnetic and gravitational-wave observations, it will be possible to estimate the DWD properties, such as mass, period, central temperature, and magnetic fields (see, e.g., \cite{2022ApJ...940...90C, 2025JHEAp..45..333N}).
Subsequently, using the mass function of these binaries, we can also constrain the rate of DWD mergers. Thus, DWD mergers open an interesting avenue for multi-wavelength and future multi-messenger investigations.

%%%%%%%%%%%%%%%%%%%%%%%%%%%%%%%%%%%%%%%%%%%%%%%%%%%%%
\section*{Acknowledgments}

This work was supported by CAPES and the Universidade Federal do Espírito Santo. R. F. M. M. thanks the University Gabriele d'Annunzio for the financial support for the Cycle XL PhD scholarship (2024--2027).
J. G. C. is grateful for the support of FAPES (1020/2022, 1081/2022, 976/2022, 332/2023), CNPq (311758/2021-5, 306018/2025-0), and FAPESP (grant No. 2021/01089-1). M. B. acknowledges the Department of Physics and Earth Science of the University of Ferrara for the financial support through the FIRD 2024 grant.

\begin{table*}[ht]
\centering
% \textcolor{red}{
\caption{Lightcurve metrics for an NS central remnant with $B_d = 10^{14}$ G and $P_0 = 1$ ms, which correspond to the maximum dipole factor examined here, $\log D = 40$. In the LSST $ugrizy$ bands, we provide the magnitude and time at peak, rise and decay times corresponding to half peak flux, horizon and detection rate upper limits as described in Section 4 of the paper.}
\label{tab:rates_NS_14_-3_LSST}
\begin{tabular}{lrrrrrrrr}
\toprule
Bands & u & g & r & i & z & y \\
$m_{\rm lim}$ (mag) & 23.9 & 25.0 & 24.7 & 24.0 & 23.3 & 22.1 \\ 
\midrule
\midrule
$M_{\rm peak}$ (mag) & -12.7 & -14.6 & -13.6 & -12.7 & -11.7 & -10.4 \\
$t_{\rm peak}$ (days) & 4.7 & 5.4 & 6.3 & 7.1 & 7.8 & 8.4 \\
$t_{\rm rise}$ (days) & 3.7 & 4.2 & 4.9 & 5.5 & 6.0 & 6.5 \\
$t_{\rm decay}$ (days) & 6.1 & 7.3 & 8.8 & 10.1 & 11.1 & 12.0 \\
$d_{\rm max}$ (Mpc) & 210 & 820 & 450 & 220 & 100 & 30 \\
$d_{\rm cad}$ ($t_{\rm cad} = 2$ days) (Mpc) & 200 & 790 & 440 & 210 & 100 & 30 \\
$d_{\rm cad}$ ($t_{\rm cad} = 3$ days) (Mpc) & 190 & 760 & 430 & 210 & 100 & 30 \\
$\mathcal{N} (d_{\rm max})$ (yr$^{-1}$) & 2e+04 & 1e+06 & 2e+05 & 3e+04 & 3e+03 & 90 \\
$\mathcal{N} (d_{\rm cad})$ ($t_{\rm cad} = 2$ days) (yr$^{-1}$)  & 6e+03 & 4e+05 & 7e+04 & 8e+03 & 7e+02 & 20 \\
$\mathcal{N} (d_{\rm cad})$ ($t_{\rm cad} = 3$ days) (yr$^{-1}$)  & 8e+03 & 5e+05 & 9e+04 & 1e+04 & 1e+03 & 40 \\
\bottomrule
\end{tabular}
\end{table*}

\begin{table*}[ht]
\centering
% \textcolor{red}{
\caption{Lightcurve metrics for an NS-powered transient in the LSST and the ZTF $g$ band for selected values of the dipole factor ($\log D$) in Figure \ref{fig:lightcurve}.}
\label{tab:rates_NS_D_g_LSST_ZTF}
\begin{tabular}{lrrrr|rrrr}
\toprule
Survey ($m_{g, \rm lim}$) & \multicolumn{4}{c|}{LSST ($25.0$ mag)} & \multicolumn{4}{c}{ZTF ($20.8$ mag)} \\
\midrule
$\log D$ & 24 & 32 & 36 & 40 & 24 & 32 & 36 & 40 \\
\midrule
\midrule
$M_{\rm peak}$ (mag) & -7.7 & -7.7 & -12.2 & -14.6 & -6.7 & -6.7 & -11.2 & -13.7 \\
$t_{\rm peak}$ (days) & 0.2 & 0.2 & 4.0 & 5.4 & 0.2 & 0.2 & 3.2 & 4.7 \\
$t_{\rm rise}$ (days) & 0.2 & 0.2 & 2.8 & 4.2 & 0.1 & 0.1 & 2.3 & 3.7 \\
$t_{\rm decay}$ (days) & 0.3 & 0.3 & 5.0 & 7.3 & 0.2 & 0.2 & 3.9 & 6.1 \\
$d_{\rm max}$ (Mpc) & 30.0 & 30.0 & 280.0 & 820.0 & 0 & 0 & 20 & 80 \\
$d_{\rm cad}$ ($t_{\rm cad} = 2$ days) (Mpc) & 0.0 & 0.0 & 260.0 & 790.0 & 0 & 0 & 20 & 80 \\
$d_{\rm cad}$ ($t_{\rm cad} = 3$ days) (Mpc) & 0.0 & 0.0 & 240.0 & 760.0 & 0 & 0 & 20 & 70 \\
$\mathcal{N} (d_{\rm max})$ (yr$^{-1}$) & 1e+02 & 1e+02 & 6e+04 & 1e+06 & 0 & 0 & 40 & 1e+03 \\
$\mathcal{N} (d_{\rm cad})$ ($t_{\rm cad} = 2$ days) (yr$^{-1}$)  & 0 & 0 & 1e+04 & 4e+05 & 0 & 0 & 10 & 3e+02 \\
$\mathcal{N} (d_{\rm cad})$ ($t_{\rm cad} = 3$ days) (yr$^{-1}$)  & 0 & 0 & 2e+04 & 5e+05 & 0 & 0 & 10 & 4e+02 \\
\bottomrule
\end{tabular}
\end{table*}

\bibliographystyle{elsarticle-num} %elsarticle-harv

\bibliography{references}

\appendix

%%%%%%%%%%%%%%%%%%%%%%%%%%%%%%%%
\section{Ejecta expansion model}
\label{app:ejecta}
%%%%%%%%%%%%%%%%%%%%%%%%%%%%%%%%

% Para goal: set the geometry of the ejecta
Following the approach of \cite{2023ApJ...958..134S}, the ejecta is discretized into $i=0,1,...,N$ shells, where $i=0$ represents the innermost shell to the star and $i=N$ the outermost. 
We adopt that each shell evolves independently with radius and velocity
\begin{equation}\label{eq:ri}
r_i(t) = \tilde{r}_i \left( \frac{t}{t_*} \right)^n\,\quad;\quad  v_i = \tilde{v}_i\left(\frac{t}{t_*}\right)^{n-1}\,,  
\end{equation}
with $t_*=n \tilde{r}_{0} / \tilde{v}_{0}$ as the characteristic expansion timescale, $\tilde{r}_i$ and $\tilde{v}_i$ are the initial position and initial velocity of the $i$-th shell, and $n$ is an expansion parameter. The ejecta density profile is given by
\begin{equation}\label{eq:rho}
    \rho(t) = \tilde\rho_0\left(\frac{\tilde{r}_0}{\tilde{r}_i}\right)^m \left(\frac{t}{t_*}\right)^{-3n}\,,
\end{equation}
where $m$ is the power-law index and $\tilde{\rho}_0$ is the normalising constant defined as
\begin{equation}\label{eq:rho0}
    \tilde{\rho}_0=\frac{3-m}{4\pi}\frac{m_{ej}}{\tilde{r}_0^3}\left[\left(\frac{\tilde{r}_N}{\tilde{r}_{0}}\right)^{3-m}-1\right]^{-1}\,.
\end{equation}
Assuming a constant gray opacity $\kappa$ throughout the ejecta, we have the optical depth of the $i$-th layer
\begin{equation}\label{eqn:tau_i}
    \tau_i(t)=\int_{r_i}^\infty\kappa \rho(r,t)dr=\tilde{\tau}_i\left(\frac{t}{t_*}\right)^{-2n}\,,
\end{equation}
with the constant $\tilde{\tau}_i$ given by 
\begin{equation}\label{eq:tauitilde}
\tilde{\tau}_{i} = \frac{m-3}{m-1} \frac{\kappa\, m_{\rm ej}}{4\pi \tilde{r}_0^2} \left[ \frac{\left(\frac{\tilde{r}_0}{r_i}\right)^{m-1} - \left(\frac{\tilde{r}_0}{r_N}\right)^{m-1}}{1 - \left(\frac{\tilde{r}_0}{\tilde{r}_N}\right)^{m-3}} \right]\,.
\end{equation}
The integration (Eq. \ref{eqn:tau_i}) from the $i$-th shell to the furthest point from the remnant, taken as the radius of the outermost layer of the ejecta, gives the optical depth to the given layer $i$ of the ejecta. 
Thus, the optical depth for the entire ejecta is $\tau_0$. 
\\The photosphere radius of the ejecta is then given as
\begin{equation}\label{eqn:photo}
    R_{\rm ph} = 
	\begin{cases}
		r_N (1 - \xi r_0^2 )^{1/(1-m)} & \text{if } \tau \ll 1,\\
		r_0 & \text{if } \tau \geq 1\,,
	\end{cases}{}
\end{equation}
where
\begin{equation}\label{eq:xi}
    \xi = \frac{4 \pi}{\kappa\,m_{\rm ej}} \frac{1-m}{m-3} \left(\frac{\tilde{r}_0}{\tilde{r}_N}\right)^{1-m} \left[1 - \left(\frac{\tilde{r}_0}{\tilde{r}_N}\right)^{m-3}\right]\,.
\end{equation}

Our model assumes a power-law density profile, a constant gray opacity, and self-similar expansion of the ejecta, with $n=1$ and $m=9$. Although this is a standard and computationally efficient approximation, future hydrodynamic simulations of DWD mergers may refine these assumptions. In particular, variations in opacity (e.g., due to chemical stratification or recombination) or deviations from spherical symmetry could affect photospheric evolution and the transient peak.

%%%%%%%%%%%%%%%%%%%%%%%%%%%%%%%%%%%%%%%%%%%%%%%%%%%%%%%%%%%%%%%%%
\section{Comparison with previous work: the case of a central WD}
\label{app:comparison}
%%%%%%%%%%%%%%%%%%%%%%%%%%%%%%%%%%%%%%%%%%%%%%%%%%%%%%%%%%%%%%%%%

\begin{figure}
    \centering
    \includegraphics[width=1\linewidth]{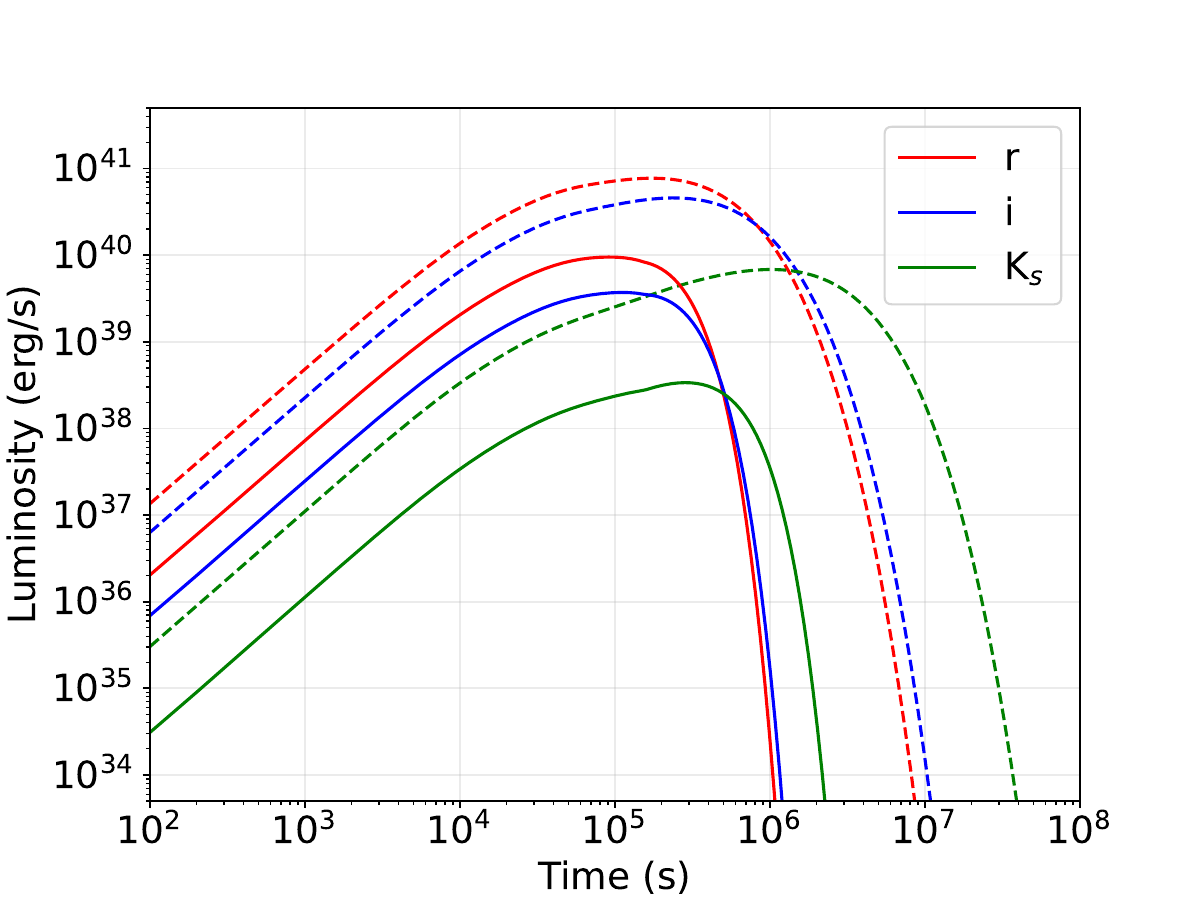}
    \caption{Comparison of the lightcurve of the optical transient powered by fallback accretion onto a central WD left by a DWD merger (solid: this work, dashed: Ref. \cite{2023ApJ...958..134S}).}
    \label{fig:WD-lightcurve}
\end{figure}

Figure \ref{fig:WD-lightcurve} presents the lightcurves of the transient produced by fallback onto the central WD, using the same model parameters considered in \cite{2023ApJ...958..134S}. One of the improvements in our work is the introduction of a leakage term, which makes the lightcurve peak earlier and fainter. With this new lightcurve and taking into consideration the effect of cadence, we obtained improved upper limits on the detection rate by LSST and ZTF presented in Table \ref{tab:rates_WD_LSST_ZTF_resub}. These values are upper limits on the detection rate, obtained from Eqs. (\ref{eq:Ndet}) and (\ref{eq:Rwdns}), with $f_{\rm rem} = 0.95$ and $k_{\rm WD} = 1$. We find that detection rates estimated using $d_{\rm max}$ are consistent, whereas the more realistic estimate, which accounts for the survey cadence, is much lower. 

\begin{table*}[ht]
\centering
% \textcolor{red}{
\caption{Lightcurve metrics for a transient powered by fallback accretion onto a central WD engine}
\label{tab:rates_WD_LSST_ZTF_resub}
\begin{tabular}{lrrr|rrr}
\toprule
Survey & \multicolumn{3}{c|}{LSST} & \multicolumn{3}{c}{ZTF} \\
\midrule
Band & g & r & i & g & r & i \\
$m_{\rm lim}$ (mag) & 25.0 & 24.7 & 24.0 & 20.8 & 20.6 & 19.9 \\
\midrule
\midrule
$M_{\rm peak}$ (mag) & -10.1 & -9.2 & -8.3 & -9.1 & -10.2 & -9.3 \\
$t_{\rm peak}$ (days) & 0.8 & 1.0 & 1.2 & 0.7 & 0.8 & 1.0 \\
$t_{\rm rise}$ (days) & 0.6 & 0.8 & 1.0 & 0.5 & 0.6 & 0.8 \\
% 0.7 & 1.0 & 1.2 & 0.6 & 0.7 & 1.0 \\
$t_{\rm decay}$ (days) & 1.5 & 1.8 & 2.1 & 1.1 & 1.5 & 1.8 \\
% 2.9 & 3.4 & 3.7 & 2.2 & 2.9 & 3.4 \\
$d_{\rm max}$ (Mpc) & 100 & 60 & 30 & 10 & 10 & 10 \\
$d_{\rm cad}$ ($t_{\rm cad} = 2$ days) (Mpc) & 60 & 40 & 20 & 0 & 10 & 0 \\
$d_{\rm cad}$ ($t_{\rm cad} = 3$ days) (Mpc) & 30 & 30 & 20 & 0 & 0 & 0 \\
$\mathcal{N} (d_{\rm max})$ (yr$^{-1}$) & 2970 & 560 & 70 & 0 & 10 & 0 \\
$\mathcal{N} (d_{\rm cad})$ ($t_{\rm cad} = 2$ days) (yr$^{-1}$)  & 150 & 50 & 10 & 0 & 0 & 0 \\
$\mathcal{N} (d_{\rm cad})$ ($t_{\rm cad} = 3$ days) (yr$^{-1}$)  & 30 & 20 & 0 & 0 & 0 & 0 \\
\bottomrule
\end{tabular}
% }
\end{table*}

%%%%%%%%%%%%%%%%%%%%%%%%%%%%%%%%%%%%%%%%%%%%%%%%%%%%%%%%%%%%%%%%%
\section{Impact of the starting time of observation on detection rates}
\label{app:impact}
%%%%%%%%%%%%%%%%%%%%%%%%%%%%%%%%%%%%%%%%%%%%%%%%%%%%%%%%%%%%%%%%%

\begin{figure}
    \centering
    \includegraphics[width=1\linewidth]{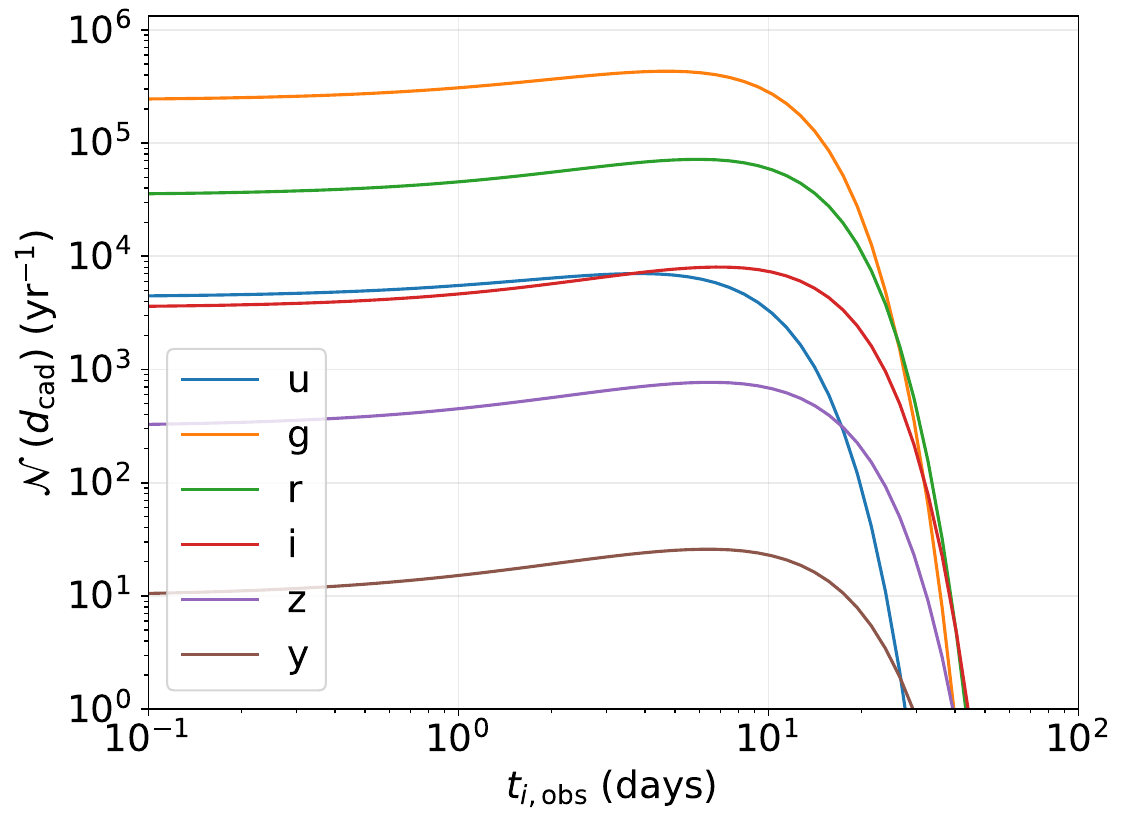}
    \caption{Impact of the time of first observation, $t_{i, \rm obs}$, on the detection rates, $\mathcal{N} (d_{\rm cad})$ (yr$^{-1}$) in LSST-$ugrizy$ filters and $t_\text{cad} = 2$ days. The detection rates correspond to the transient from an NS central remnant with  $\log D = 40$.}
    \label{fig:discovery-impact}
\end{figure}

For a fast, faint transient source, the first observation significantly affects the detection and characterisation of the transient. In Tables \ref{tab:rates_NS_14_-3_LSST}, \ref{tab:rates_NS_D_g_LSST_ZTF}, and \ref{tab:rates_WD_LSST_ZTF_resub}, we have assumed the observation of the source starts at the time of the peak of the lightcurve, i.e., $t_{i, \rm obs} = t_{\rm peak}$. In practice, observations can begin during the rising or the declining phase of the light curve. In this appendix, we quantify the effect of the value $t_{i, \rm obs}$ on the detection rate. 

Figure \ref{fig:discovery-impact} shows the detection rate as a function of $t_{i, \rm obs}$ for a cadence of 2 days, for the case of an NS $\log D = 40$. The detection rate for $t_{i, \rm obs} = t_{\rm peak}$ can be found in Table \ref{tab:rates_NS_14_-3_LSST}. We have checked that the time $t_{i, \rm obs}$ at which the detection rate is the highest corresponds to the peak time of the lightcurve. Thus, we obtain an upper limit on the detection rate with at least two observations when the first observation is at the light-curve peak. When $t_{i, \rm obs} < t_{\rm peak}$, the detection rate decreases less than one order of magnitude relative to the maximum, while it can be drastically reduced when $t_{i, \rm obs} > t_{\rm peak}$.

\end{document}